\documentclass[prc,superscriptaddress,unsortedaddress,twocolumn,showpacs]{revtex4}
\usepackage{amsmath}
\usepackage{amssymb}
\usepackage{times}
\usepackage{mathrsfs}
\usepackage{multirow}
\usepackage{bm}
\usepackage{wrapft}
\usepackage{color}
\usepackage{graphicx}
\usepackage{ulem}

\def\fun#1#2{\lower3.6pt\vbox{\baselineskip0pt\lineskip.9pt
\ialign{$\mathsurround=0pt#1\hfil##\hfil$\crcr#2\crcr\sim\crcr}}}

\newcommand{\beq}{\begin{equation}}
\newcommand{\eeq}{\end{equation}}
\newcommand{\bea}{\begin{eqnarray}}
\newcommand{\eea}{\end{eqnarray}}

\newcommand{\bfi}[1]{\mbox{\boldmath $#1$}}

\newcommand{\vp}{{\bfi p}}

\newcommand{\vs}{{\bfi s}}

\newcommand{\vt}{{\bfi t}}

\renewcommand{\vr}{{\bfi r}}
\newcommand{\vR}{{\bfi R}}

\begin{document}

\title{
Microscopic coupled-channel calculations of nucleus-nucleus scattering including chiral three-nucleon-force effects
}

\author{Kosho Minomo}
\email[]{minomo@rcnp.osaka-u.ac.jp}
\affiliation{Research Center for Nuclear Physics, Osaka University, Ibaraki 567-0047, Japan}

\author{Michio Kohno}
\affiliation{Research Center for Nuclear Physics, Osaka University, Ibaraki 567-0047, Japan}

\author{Kazuyuki Ogata}
\affiliation{Research Center for Nuclear Physics, Osaka University, Ibaraki 567-0047, Japan}

\date{\today}

\begin{abstract}
We analyze $^{16}$O-$^{16}$O and $^{12}$C-$^{12}$C scattering with the microscopic coupled-channels method
and investigate the coupled-channels and three-nucleon-force (3NF) effects on elastic and inelastic cross sections.
In the microscopic coupled-channels calculation, the Melbourne $g$-matrix interaction modified
according to the chiral 3NF effects is used.
It is found that the coupled-channels and 3NF effects additively change both the elastic and inelastic cross sections.
As a result, the coupled-channels calculation including the 3NF effects significantly
improves the agreement between the theoretical results and the experimental data.
The incident-energy dependence of the coupled-channels and 3NF effects is also discussed.
\end{abstract}


\maketitle

{\it Introduction.}
Microscopic description of many-body nuclear reactions is a long standing subject in nuclear physics.
Elastic scattering, which is one of the most basic processes, was first understood phenomenologically with the optical model.
The model was founded by the multiple scattering theory~\cite{Wat58,Ker59,Yah08} from the microscopic point of view.
According to the theory, elastic scattering is described by multistep processes with a nucleon-nucleon effective interaction.
So far many types of density-dependent complex interactions ($g$-matrix interactions)
have been proposed as effective interactions based on the Brueckner theory.
It is, however, still not easy to reproduce even the elastic scattering data microscopically.

Recently, the Melbourne group achieved a great success to describe proton scattering with no adjustable parameter.
They applied a $g$-matrix interaction constructed with the Bonn potential~\cite{Mac87} to the folding model calculation,
and reproduced the measured elastic cross sections and analyzing powers in a wide incident-energy range~\cite{Amo00}.
In addition, the Melbourne interaction was also applied to neutron scattering~\cite{Amo02} and
$^6$He-proton scattering~\cite{Kar13,Toy13}.
Furthermore, total reaction cross sections of nucleus-nucleus scattering were well reproduced~\cite{Min12,Wat14}.
These are the remarkable progress of microscopic reaction theory in recent years.
In some cases, however, nucleus-nucleus elastic cross sections cannot be reproduced~\cite{Min14}.
One of the reasons for this failure may be three-nucleon force (3NF) effects on nucleus-nucleus scattering;
the Melbourne interaction does not include the effects explicitly.

3NF effects on many-body nuclear reactions are a hot topic relating to
the equation-of-state of nuclear matter.
In the $g$-matrix folding model, the 3NF effects are represented through
the density dependence of $g$-matrix.
In Refs.~\cite{Fur08,Fur09}, the repulsive nature due to 3NF effects
is simulated by changing the two-nucleon force (2NF) in the nuclear medium to reproduce the empirical saturation.
The $g$-matrix interaction (CEG07) obtained by including the \lq\lq3NF'' effects gives
a less attractive and less absorptive potential than the original one.
The effects are rather small for proton scattering
but the agreement between the theoretical results and the experimental values
are improved for the spin observables at intermediate energies~\cite{Fur08}.
For nucleus-nucleus scattering, the measured elastic cross sections
even at large scattering angles are well reproduced~\cite{Fur09}.
In Ref.~\cite{Raf13}, as a first attempt of explicitly applying the phenomenological 3NF to nucleon optical potential,
the Urbana 3NF~\cite{Pud97} and the three-nucleon interaction~\cite{Fri81,Lag81} were used to construct $g$-matrix interactions.
The $g$-matrices were applied to proton elastic scattering, which showed the 3NF effects are again small.

In the previous works~\cite{Min14,Toy14},
the roles of 3NFs based on the chiral effective field theory~\cite{Epe05,Epe09} were investigated
in nucleon-nucleus and nucleus-nucleus elastic scattering.
One of the advantages of using the chiral interactions~\cite{Epe05,Epe09,Heb11,Mac11} is that
the two-, three-, and many-body forces are treated systematically
so that the uncertainty of interactions is minimized.
At present, the chiral interactions are most reliable for investigating the 3NF effects
although they cannot be applied to high-energy reactions beyond the cutoff energy.
In Refs.~\cite{Min14,Toy14},
the Melbourne interaction was modified according to the effects of the chiral NNLO 3NF~\cite{Epe05,Epe09},
based on the knowledge that $g$-matrices from the Bonn potential and the chiral 2NF
are resembling; see Refs.~\cite{Min14,Toy14} for details.
The chiral 3NF effects make the real (imaginary) part of the folding potential
less attractive (much absorptive).
This provides us a new insight into the 3NF effects, which had not been
acquired in the studies employing phenomenological 3NFs.
As a result, the chiral 3NF effects improve the agreement between the theoretical calculation and the experimental data
for nucleus-nucleus elastic scattering.
The recent study by Toyokawa {\it et al}.~\cite{Toy15} in which g-matrices are constructed directly
using the chiral 2NFs and 3NFs reinforces previous findings.

Although the 3NF effects improve the description of nucleus-nucleus elastic scattering, there remain some discrepancies.
The theoretical cross sections at large angles overshoot the data for $^{16}$O-$^{16}$O scattering,
whereas the theoretical result is slightly off the data in the diffraction pattern of $^{12}$C-$^{12}$C scattering.
One of the reasons for this discrepancy will be the coupled-channels effects due to the projectile and target excitations.
In fact, these effects have been discussed in the preceding studies~\cite{Kho08,Tak10} based on the microscopic coupled-channels method.
In the analyses inelastic cross sections were mainly focused and
elastic cross sections were used just to determine the real and/or imaginary parts of the optical potentials.
On the other hand, our interest in this report is to clarify the coupled-channels effects on both the
elastic and inelastic nucleus-nucleus scattering.

For this purpose, we will perform microscopic coupled-channels calculations,
without any adjustable parameters, of $^{16}$O-$^{16}$O and $^{12}$C-$^{12}$C scattering
using the Melbourne interaction modified with the chiral 3NF effects, and
clarify the coupled-channels and 3NF effects on scattering observables.
These two effects are treated simultaneously within the microscopic coupled-channels framework,
and their incident-energy dependence is also investigated.

{\it Framework of coupled-channels calculation.}
Since the microscopic coupled-channels method to describe
elastic and inelastic scattering is well established, we give just a brief review of it;
see, e.g., Refs.~\cite{Kho00,Sak88,Kat02} for details.
We consider scattering of a projectile P off a target T, where P and T are identical bosons.
The radial part $\chi_{\gamma L,\gamma_0^{}L_0}^{J}(R)$ of the partial wave
between P and T, where $L$ is the P-T relative orbital angular momentum and $J$ is the total angular momentum,
is obtained by solving the following coupled-channels equations,
\begin{widetext}
\bea
&&
\bigg[-\frac{\hbar_{}^2}{2\mu}\frac{d^2}{dR^2}
+\frac{\hbar_{}^2}{2\mu}\frac{L(L+1)}{R_{}^2}
+F_{\gamma L,\gamma L}^{J}(R)
+U^{\rm Coul}(R)-E_\gamma^{}\bigg]
\chi_{\gamma L,\gamma_0^{}L_0}^{J}(R)
=
-\sum_{\gamma'L'\neq \gamma L}^{}F_{\gamma L,\gamma'L'}^{J}(R)
\chi_{\gamma'L',\gamma_0^{}L_0}^{J}(R)
\nonumber\\
\label{cceq}
\eea
\bea
F_{\gamma L,\gamma'L'}^{J}(R)
&=&
\sum_{\lambda}
i^{L'-L}
(-1)^{L'-L+S-J+\lambda}
\sqrt{(2L+1)(2L'+1)}
W(LSL'S'|J\lambda)(L'0L0|\lambda 0)
\nonumber\\
&&\times
2N_{I_{\rm P}^{}I_{\rm T}^{}}N_{I'_{\rm P}I'_{\rm T}}
\Big(U_{\gamma(SI_{\rm P}^{}I_{\rm T}^{}),\gamma'(S'I'_{\rm P}I'_{\rm T})}^{\lambda}(R)
+(-1)^S U_{\gamma(SI_{\rm T}^{}I_{\rm P}^{}),\gamma'(S'I'_{\rm P}I'_{\rm T})}^{\lambda}(R)\Big),
\eea
\bea
&&U_{\gamma(SI_{\rm P}^{}I_{\rm T}^{}),\gamma'(S'I'_{\rm P}I'_{\rm T})}^{\lambda}(R)
\nonumber\\
&=&
\frac{1}{\sqrt{4\pi}}
\sqrt{(2S+1)(2S'+1)(2I_{\rm P}^{}+1)(2I_{\rm T}^{}+1)}
\sum_{\lambda_{\rm P}^{}\lambda_{\rm T}^{}}
i^{\lambda_{\rm P}^{}+\lambda_{\rm T}^{}}
\left\{
\begin{array}{ccc}
\displaystyle
 I_{\rm P}^{} & I_{\rm T}^{} & S \\
 I'_{\rm P} & I'_{\rm T} & S' \\
 \lambda_{\rm P}^{} & \lambda_{\rm T}^{} & \lambda \\
\end{array}
\right\}
\nonumber\\
&&\times
\bigg\{
\int\!d\hat{\vR}~d\vr_{\rm P}^{}~d\vr_{\rm T}^{}~
\rho_{I_{\rm P}^{}I'_{\rm P}}^{\lambda_{\rm P}^{}}(r_{\rm P}^{})
\rho_{I_{\rm T}^{}I'_{\rm T}}^{\lambda_{\rm T}^{}}(r_{\rm T}^{})
g^{\rm (dr)}(s;\rho)
\bigg[Y_{\lambda}^{}(\hat{\vR})\otimes
\big[Y_{\lambda_{\rm P}^{}}^{}(\hat{\vr}_{\rm P}^{})
\otimes
Y_{\lambda_{\rm T}^{}}^{}(\hat{\vr}_{\rm T}^{})
\big]_{\lambda}^{}
\bigg]_{00}
\nonumber\\
&&+
\int\!d\hat{\vR}~d\vp~d\vs~
\rho_{I_{\rm P}^{}I'_{\rm P}}^{\lambda_{\rm P}^{}}(r_{\rm P}^{})
\hat{j}_1^{}\big(k_{\rm F}^{}(p)s\big)
\rho_{I_{\rm T}^{}I'_{\rm T}}^{\lambda_{\rm T}^{}}(r_{\rm T}^{})
\hat{j}_1^{}\big(k_{\rm F}^{}(t)s\big)
g^{\rm (ex)}(s;\rho)
j_0^{}\big(k(R)s/M\big)
\bigg[Y_{\lambda}^{}(\hat{\vR})\otimes
\big[Y_{\lambda_{\rm P}^{}}^{}(\hat{\vp})
\otimes
Y_{\lambda_{\rm T}^{}}^{}(\hat{\vt})
\big]_{\lambda}^{}
\bigg]_{00} \bigg\}.
\nonumber\\
\label{couppot}
\eea
\end{widetext}
Here, $\vs=\vR+\vr_{\rm P}^{}-\vr_{\rm T}^{}$, $\vp=\vr_{\rm P}^{}-\vs/2$, and $\vt=\vr_{\rm T}^{}+\vs/2$.
The channel number indicating the intrinsic spins of P ($I_{\rm P}^{}$) and T ($I_{\rm T}^{}$) is specified by $\gamma$,
and the channel spin $S$ defined by $|I_{\rm P}^{}-I_{\rm T}^{}|<S<I_{\rm P}^{}+I_{\rm T}^{}$.
The subscript 0 indicates the initial channel.
$E_\gamma^{}$ is defined by $E_\gamma^{}=E-\varepsilon_\gamma^{}$
with $E$ the incident energy in the center-of-mass system and
$\varepsilon_\gamma^{}$ the sum of the excitation energies of P and T.
$M$ is defined as $M=A_{\rm P}A_{\rm T}/(A_{\rm P}+A_{\rm T})$,
where $A_{\rm P(T)}$ is the mass number of P (T).
$U^{\rm Coul}(R)$ is the Coulomb potential between P and T.
$W(LSL'S'|J\lambda)$ is the Racah coefficient
and $N_{I_{\rm P}^{}I_{\rm T}^{}}
=[2(1+\delta_{I_{\rm P}^{}I_{\rm T}^{}}^{}\delta_{\varepsilon_{\rm P}^{}\varepsilon_{\rm T}^{}}^{})]^{-1/2}$.

The input of the microscopic coupled-channels calculation is
the nucleon-nucleon effective interaction $g^{\rm (dr)}$ and $g^{\rm (ex)}$,
and the transition densities $\rho_{I_{\rm A}^{}I'_{\rm A}}^{\lambda_{\rm A}^{}}$
(${\rm A}={\rm P}$ and T).
In this paper, we use the Melbourne $g$-matrix interaction modified with the chiral 3NF effects~\cite{Min14,Toy14}.
Although a $g$ matrix contains an effect of single-particle excitation in nuclear matter,
collective excitation for a specific nucleus is not adequately included.
Following preceding studies, therefore, we take into account these collective excited states using
the coupled-channels framework. A possible double-counting of the coupling to non-elastic channels
is expected to be negligible, because the channels explicitly included are 
specifically collective and have no analogue in nuclear matter.

We take the frozen density approximation for evaluating $\rho$ in the argument of the $g$ matrix;
$\rho=\rho_{\rm P}^{}(r_{\rm m})+\rho_{\rm T}^{}(r_{\rm m})$ is used, where
$r_{\rm m}$ means the midpoint of the interacting two nucleons.
For the coupling potentials,
we take the average of the densities in the initial and final states, i.e.,
$\rho_{\rm A}^{}=
(\rho_{I'_{\rm A}I'_{\rm A}}^{\lambda_{\rm A}^{}=0}
+\rho_{I_{\rm A}I_{\rm A}}^{\lambda_{\rm A}^{}=0})/2$
for ${\rm A}={\rm P}$ or T.

Since we need not only the ground state density but also transition densities for excited states,
we adopt microscopic cluster models to obtain them.
For $^{12}$C, we consider the $0_1^+$, $2_1^+$ (4.44~MeV), $0_2^+$ (7.65~MeV),
and $2_2^+$ (10.3~MeV) states, and
we use the transition densities obtained by the resonating group method (RGM)
based on a $3\alpha$ model~\cite{Kam81}.
These densities are found to reproduce the elastic and inelastic form factors
determined by electron scattering and thus highly reliable.
For $^{16}$O,
the $0_1^+$, $3_1^-$ (6.13 MeV), and $2_1^+$ (6.92 MeV) states are taken into account.
We use the transition densities obtained by the orthogonality condition model (OCM)
assuming an $\alpha+^{12}{\rm C}$ structure~\cite{Oka95}.
However, the transition densities between the excited states are not prepared; we thus neglect
the coupling between the excited states in the calculation of the $^{16}$O-$^{16}$O scattering.
It is noted that we multiply the $^{16}$O transition density between the $0_1^+$ and $3_1^-$ ($2_1^+$) states by $0.816$ ($1.17$),
so as to reproduce the experimental values of $B({\rm E}3)$ and $B({\rm E}2)$, i.e.,
$1490 \pm 70~{\rm fm}^6$ and $39.3 \pm 1.6~{\rm fm}^4$~\cite{Mis75}, respectively.
A similar procedure was adopted in Refs.~\cite{Kat02,Tak09a} using the same transition densities of $^{16}$O.
In the reaction calculation, we do not use the excitation energies of $^{12}$C and $^{16}$O
evaluated by the microscopic models but adopt the experimental values of them.

In the present calculation, we treat transitions only by the nuclear interaction,
though the Coulomb interaction is included in the diagonal components of the coupling potentials.
The symmetrization of the identical bosons as well as the mutual excitation is included properly.
We adopt the relativistic kinematics of P and T.
As for the integration for calculating the coupling potentials in Eq.~\eqref{couppot},
we perform the Monte-Carlo integration with random numbers generated by the Mersenne-Twister method~\cite{Mat98}.

\begin{figure}[tbp]
\begin{center}
\includegraphics[width=0.45\textwidth,clip]{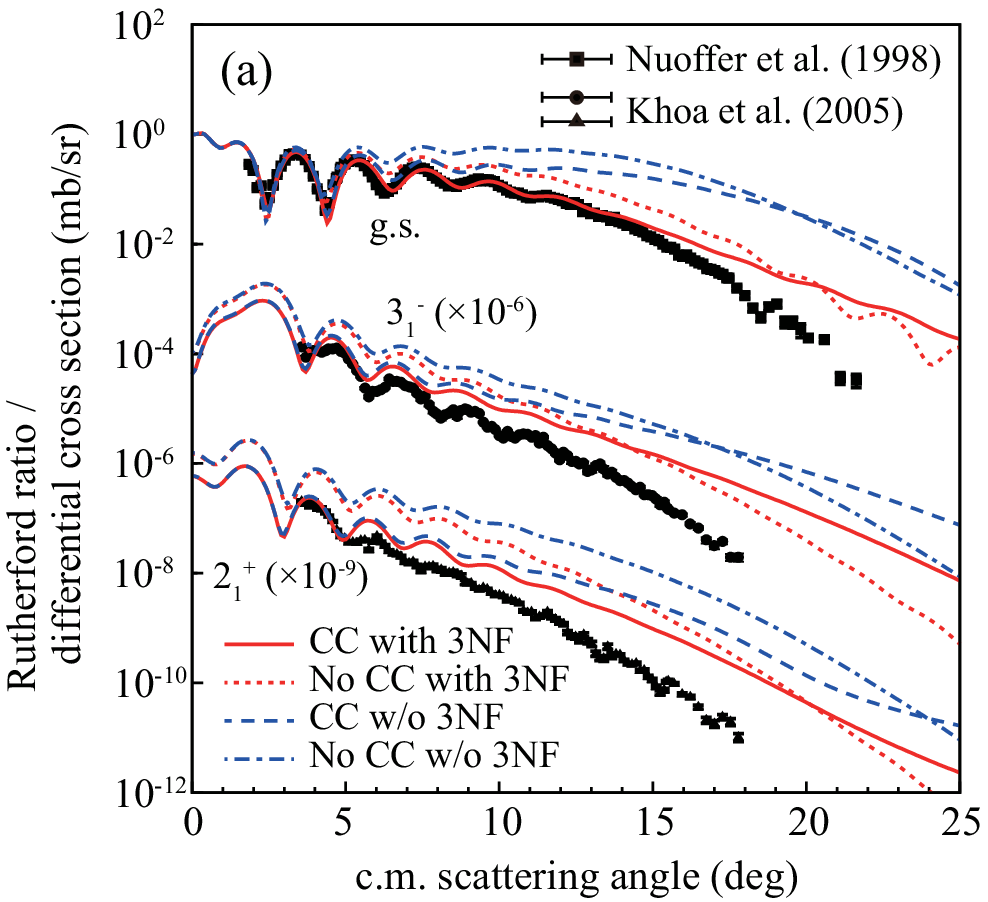}
\includegraphics[width=0.45\textwidth,clip]{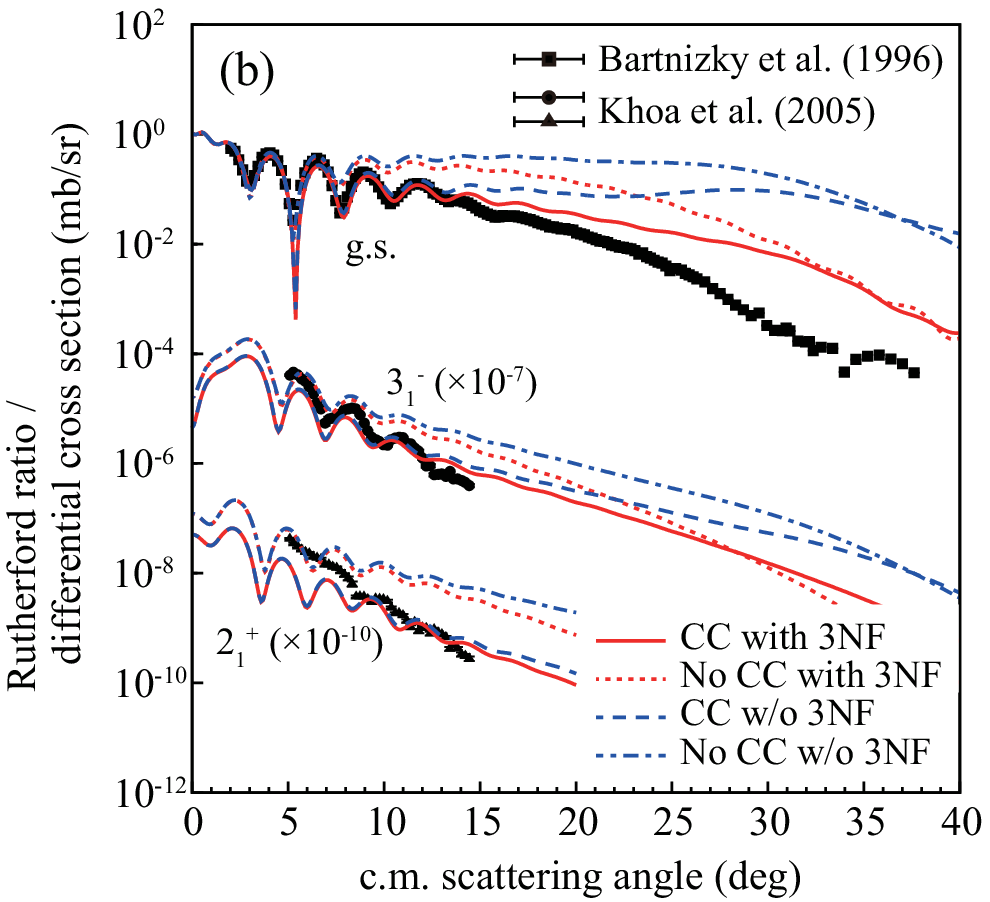}
\caption{(Color online)
Differential cross sections of elastic and inelastic
$^{16}$O-$^{16}$O scattering at (a) 70 MeV/nucleon and (b) 44 MeV/nucleon.
The solid (dashed) line corresponds to the result with (without) the 3NF effects, and
the dotted (dot-dashed) line corresponds to the result of the one-step
calculation with (without) the 3NF effects.
The experimental data are taken from Refs.~\cite{Nuo98,Kho05,Bar96}.
The inelastic cross sections are scaled by the factor shown inside the
panel for clarity.
}
\label{fig1}
\end{center}
\end{figure}

{\it Results and discussions.}
Figure~\ref{fig1} shows the differential cross sections for the $^{16}$O-$^{16}$O scattering
at (a) 70MeV/nucleon and (b) 44MeV/nucleon as a function of the scattering angle $\theta$
in the center-of-mass system.
In each panel, three cross sections
corresponding to the $0_1^+$ (top), $3_1^-$ (middle), and $2_1^+$ (bottom) states of the ejectile
are shown; the other particle is in the ground state in the final state.
For the $0_1^+$ state, i.e., the elastic scattering, the ratio to the Rutherford cross section is plotted.
The solid (dashed) line is the result of the coupled-channels calculation with (without) the 3NF effects, and
the dotted (dot-dashed) line corresponds to the result of the one-step calculation
with (without) the 3NF effects.
The experimental data are taken from Refs.~\cite{Nuo98,Kho05,Bar96}.

From Fig.~1(a), we see that the calculation including both the 3NF and coupled-channels effects
well reproduces the elastic scattering data up to $\theta=15^\circ$.
The 3NF and coupled-channels effects are found to be comparable at forward angles $\theta<12^\circ$,
whereas the former is more important than the latter
at larger angles.
As shown in Ref.~\cite{Min14}, the 3NF effects change mainly the interior part of the
nucleus-nucleus potential, which is influential to the scattering at larger angles.
On the other hand, it is well known that in general the coupled-channels
effects generate the so-called dynamical polarization potential in the nuclear surface region.
The separation of the regions sensitive to the 3NF and coupled-channels effects makes
the two effects rather independent. In fact, one sees that the 3NF (coupled-channels) effects do not
change essentially whether the coupled-channels (3NF) effects are included or not.

For the $3_1^-$ and $2_1^+$ inelastic cross sections,
the solid line reproduces the data for $\theta<5^\circ$ but slight overshooting is seen
at larger angles.
One sees that
the 3NF effects improve the agreement with the inelastic cross section data.
The 3NF effects on the inelastic cross sections appear at large angles as
in the elastic cross sections, although the effects are relatively small.
In general, inelastic cross sections at very forward angles are mainly determined by the strength of the coupling potential.
The negligibly small difference between the solid and dashed lines at forward angles
suggests the small 3NF effects on the coupling potentials.
It is found that the 3NF effects on the inelastic cross sections come from
the changes in the diagonal part of the $^{16}$O-$^{16}$O potential for each channel.
It was pointed out in Ref.~\cite{Tak09a} that the inelastic cross sections may be varied
by couplings with even higher excited states.
This can be a reason for the discrepancy for the $3_1^-$ and $2_1^+$ cross sections
at larger angles and should be investigated in the future.

At 44 MeV/nucleon (Fig.~1(b)),
the coupled-channels effects are larger and the 3NF effects are smaller than at 70 MeV/nucleon.
This is because that at lower energies the reaction takes place in the peripheral region.
Nevertheless, the 3NF effects are still non-negligible at large scattering angles.
To be precise, the correspondence between the calculation and the data is worsened from at 70~MeV/nucleon.
One of the reasons for this will be the fact that at lower incident energies the
higher excited states, which are not taken into account, may play a more significant role.

\begin{figure}[tbp]
\begin{center}
\includegraphics[width=0.45\textwidth,clip]{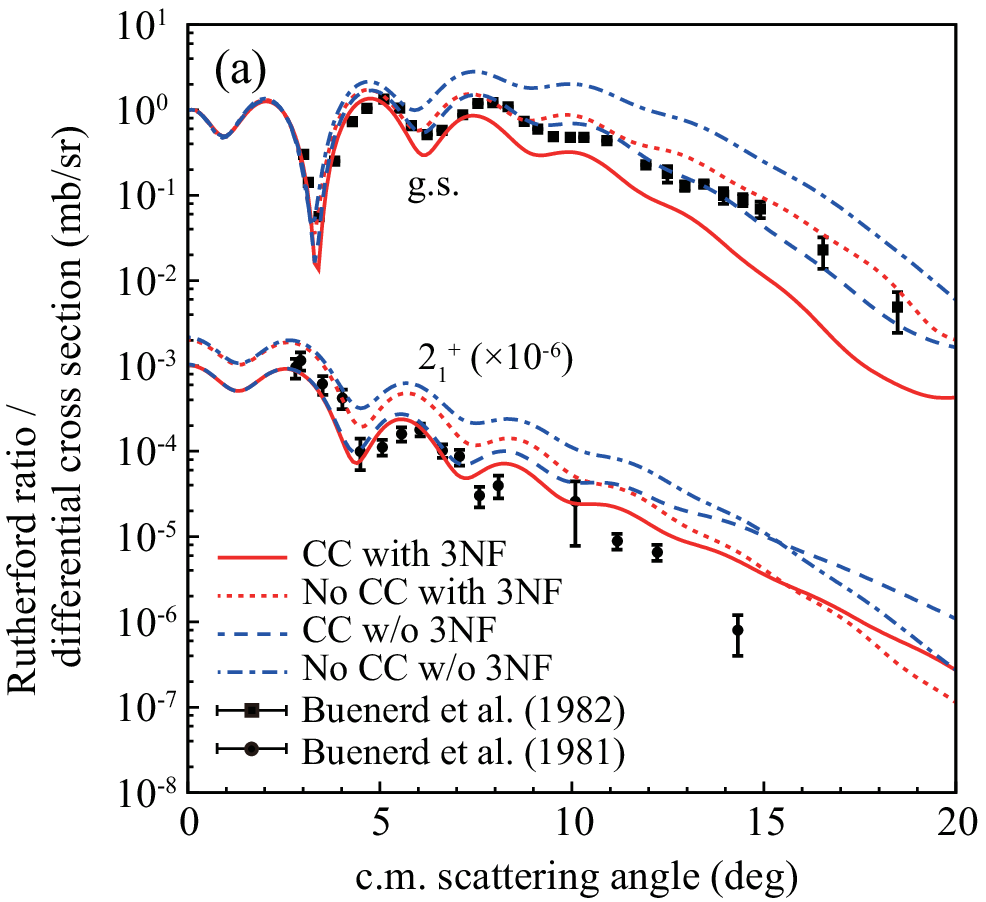}
\includegraphics[width=0.45\textwidth,clip]{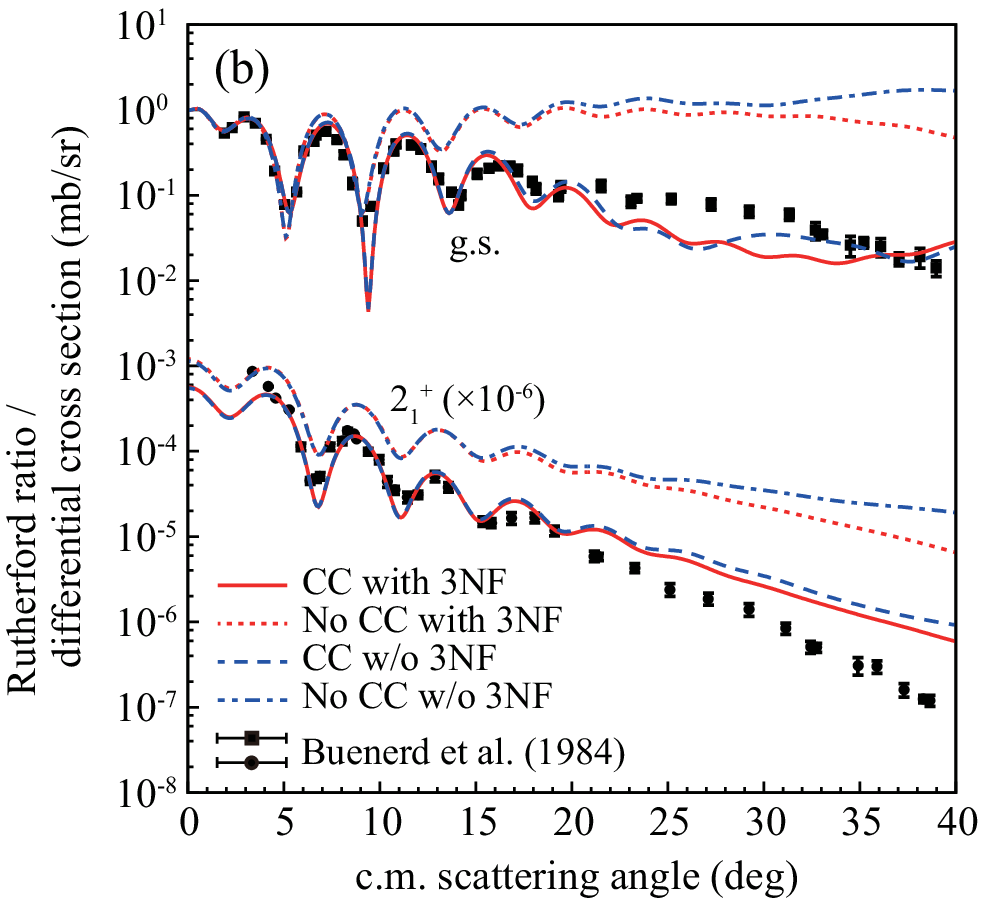}
\caption{(Color online)
Differential cross sections of elastic and inelastic
$^{12}$C-$^{12}$C scattering at (a) 85 MeV/nucleon and (b) 30 MeV/nucleon.
The meaning of the lines is the same as in Fig.~\ref{fig1}.
The experimental data are taken from Refs.~\cite{Bue82,Bue81,Bue84}.
}
\label{fig2}
\end{center}
\end{figure}

Figure~2 shows the differential cross sections for the $^{12}$C-$^{12}$C scattering
at (a) 85~MeV/nucleon and (b) 30~MeV/nucleon.
The meaning of the lines is the same as in Fig.~\ref{fig1}.
We show only the elastic and the $2_1^+$ inelastic cross section, for which the experimental data are available.
It should be noted that the $0_1^+$, $2_1^+$, $0_2^+$, and $2_2^+$ states of $^{12}$C
are taken into account in the calculation as mentioned before.
The experimental data are taken from Refs.~\cite{Bue82,Bue81,Bue84}.

At 85~MeV/nucleon, we find larger coupled-channels effects
than those in the $^{16}$O-$^{16}$O scattering.
As a result, the solid line decreases significantly and undershoots the data.
It is found that the coupling between the $0_1^+$ and $2_1^+$ states plays
a dominant role in the $^{12}$C-$^{12}$C scattering.
Unfortunately, coupled-channels effects give no change in the diffraction pattern.
For the inelastic cross section the solid line well agrees with the data.
At 30~MeV/nucleon, as shown in Fig.~2(b),
the 3NF effects do not affect to both the elastic and inelastic cross sections.
The measured elastic and inelastic cross sections are well reproduced by the coupled-channels calculations,
except for the small difference at large scattering angles.

In Table~\ref{tab1}, we show the total reaction cross sections $\sigma_{\rm R}$ obtained
by the coupled-channels and one-step calculations; the 3NF effects are included in both calculations.
One sees that $\sigma_{\rm R}$ increases by 2--4\%
depending on the strength of the coupled-channels effects.
For the $^{12}$C-$^{12}$C scattering, the calculated results are slightly larger than
the experimental values~\cite{Tak09b,Kox84}.
This may be due to the fact that the root-mean-square (RMS) radius of the proton density calculated by the RGM
is larger than its empirical value by 3~\%.
Since $\sigma_{\rm R}$ is very sensitive to the proton and neutron RMS radii,
some fine tuning of the microscopic density will be needed for a detailed analysis of $\sigma_{\rm R}$.

\begin{table}[bth]
\caption{
Calculated and measured total reaction cross sections $\sigma_{\rm R}$ (in the unit of mb).
The results calculated with and without coupled-channels effects are shown;
the 3NF effects are taken into account in both calculations.
}
\begin{center}
\begin{tabular}{c|cccc} \hline\hline
system & & CC & 1ch & Exp. \\ \hline
$^{16}$O-$^{16}$O@70MeV/nucleon & & $1453$ & $1426$ & -- \\
$^{16}$O-$^{16}$O@44MeV/nucleon & & $1519$ & $1473$ & -- \\
$^{12}$C-$^{12}$C@85MeV/nucleon & & $1078$ & $1052$ &
$998 \pm 13$\footnote{At 86.3 MeV/nucleon~\cite{Tak09b}},
$960 \pm 30$\footnote{At 83 MeV/nucleon~\cite{Kox84}} \\
$^{12}$C-$^{12}$C@30MeV/nucleon & & $1262$ & $1214$ &
$1209 \pm 32$\footnote{At 32.5 MeV/nucleon~\cite{Tak09b}},
$1315 \pm 40$\footnote{At 30 MeV/nucleon~\cite{Kox84}} \\ \hline\hline
\end{tabular}
\label{tab1}
\end{center}
\end{table}

{\it Summary.}
We have analyzed the elastic and the $3_1^-$ and $2_1^+$ inelastic cross sections
for the $^{16}$O-$^{16}$O scattering at 70 and 44~MeV/nucleon,
and also the elastic and the $2_1^+$ inelastic cross section
for the $^{12}$C-$^{12}$C scattering at 85 and 30 MeV/nucleon, by means of
the microscopic coupled-channels method.
The coupled-channels and 3NF effects on these cross sections were investigated.

Since the coupled-channels and 3NF effects additively change both the elastic and inelastic cross sections,
it is important to consider the two effects together for a quantitative
comparison between the calculation and the experimental data for nucleus-nucleus scattering.
It was found that the 3NF effects on the inelastic cross sections are almost the same as on the elastic cross sections.

In detail, the relative importance of the 3NF and coupled-channels effects depends on the incident energies and
scattering angles.
A key feature for understanding this dependence is the fact
that the 3NF effects strongly affect the interior part of the nucleus-nucleus potential,
whereas the coupled-channels effects affect the surface part of it.
As a general tendency, we can say that the 3NF (coupled-channels) effects are important
at relatively high (low) incident energies and at large (small) scattering angles.
It was found that the 3NF effects are negligibly small at lower than about 30 MeV/nucleon.

We have also shown that the 3NF effects on the coupling potentials
are quite small.
This may allow one to neglect the 3NF effects in the analysis of the inelastic cross sections
at very forward angles.
The coupled-channels effects are found to increase the total reaction cross sections by about a few percent.

In conclusion,
the coupled-channels calculation including the 3NF effects significantly
improves the agreement between the theoretical results and the experimental data.
However, there remain some discrepancies between them, at backward angles in particular.
It will be important to perform coupled-channels calculations including even higher excited states.
A more detailed examination of the effective interactions and transition densities
will be another important subject.

{\it Acknowledgements.}
The authors thank M. Kamimura for providing the $^{12}$C RGM densities.
The authors also thank S. Okabe, T. Furumoto, Y. Sakuragi
for their help of using the $^{16}$O OCM densities.
The numerical calculations in this work were performed at RCNP.
This work is supported in part by
Grant-in-Aid for Scientific Research
(Nos. 25400255 and 25400266)
from the Japan Society for the Promotion of Science (JSPS)
and by the ImPACT Program of Council for Science,
Technology and Innovation (Cabinet Office, Government of Japan).


\end{document}